\documentclass[12pt]{iopart}
\usepackage{graphicx} %[pdftex]
\usepackage{array}
\usepackage{SIunits}
\usepackage[indent=0cm]{caption}
\usepackage{floatflt}
\usepackage{color} % %
\def\T2{$T_2$}
\def\Pr3{$\mathrm{Pr^{3+}}$}
\def\ket#1{$\left|#1\right>$}
\def\mket#1{\left|#1\right>}

\def\PrYSO{Pr$^{3+}$:Y$_2$SiO$_5\,$}
\def\mrm#1{$\mathrm{#1}$}

\def\eq#1#2{\begin{equation}\label{#1}#2\end{equation}}
\def\al{$\alpha L\,$}

\begin{document}

\title{Extracting high fidelity quantum computer hardware from random systems}

\author{A.~Walther, B.~Julsgaard\footnote{Now with Department of Physics and Astronomy, Aarhus University, Ny Munkegade 120, DK-8000, Aarhus C}, L.~Rippe\footnote{Now with National Center for Atmospheric Research, NCAR, Boulder, CO}, Yan~Ying, S.~Kr\"{o}ll}
\address{Department of Physics, Lund University, P.O.~Box 118, SE-22100 Lund, Sweden}
\author{R.~Fisher, S.~Glaser}
\address{Department Chemie, Technische Universit\"{a}t M\"{u}nchen, Lichtenbergstrasse 4, 85747 Garching, Germany}

\ead{stefan.kroll@fysik.lth.se}

\date{\today}

\begin{abstract}An overview of current status and prospects of the development of quantum computer hardware based on inorganic crystals doped with rare earth ions is presented. Major parts of the experimental work in this area has been done in two places, Canberra, Australia and Lund, Sweden, and the present description follows more closely the Lund work. Techniques will be described that include optimal filtering of the initially inhomogeneously broadened profile down to well separated and narrow ensembles, as well as the use of advanced pulse-shaping in order to achieve robust arbitrary single-qubit operations with fidelities above 90\%, as characterized by quantum state tomography. It is expected that full scalability of these systems will require the ability to determine the state of single rare-earth ions. It has been proposed that this can be done using special readout ions doped into the crystal and an update is given on the work to find and characterize such ions. Finally, a few aspects on possibilities for remote entanglement of ions in separate rare-earth-ion-doped crystals are considered.
\end{abstract}


%\maketitle

\section{Introduction}
Many solid state quantum computing systems rely on a top down approach with carefully crafted systems, where parameters are designed during the manufacturing or growth process, for instance, superconducting qubits \cite{Nakamura1999,Plantenberg2007}, quantum dots \cite{Loss1998} and the Kane model \cite{Kane1998}. At the other end of the scale there are bottom up approaches, where useful hardware can be extracted out of random materials which intrinsically have very favorable properties, such as nitrogen vacancy (NV)-centers \cite{Jelezko2004} and rare-earth-ion-doped inorganic crystals \cite{Ohlsson2002}. In this paper we will discuss development of quantum computer hardware based on inorganic crystals doped with rare earth ions. Such crystals are used extensively in current technology, e.g. as active laser materials or as scintillating materials in X-ray detectors. They are manufactured in single crystal growth processes where a dopant material is added during the growth such that the dopant ions replaces host ions in the regular crystal lattice at concentrations that may range between 0.01\% and 10\% depending on application. Although the average concentration is defined by the dopant concentration in the crystal melt, the relative positions of the dopant ions is largely random. However, randomness and less precise control of construction and growth parameters can be compensated for by using suitable quantum computing schemes and by developing robust pulses enabling high fidelity qubit operations.

The next section (Section~\ref{RareEarth}) will describe basic properties of rare-earth-ion-doped crystals relevant for quantum information. This includes coherence times which can be several seconds \cite{Fraval2005} and ion-ion interactions as a mechanism for gate operations. For a laser beam passing through a rare-earth-ion-doped crystal there may be $~10^{15}$ ions within the laser focal region. However, qubits can still be created and prepared in well defined states and arbitrary qubit operations, characterized by full quantum state tomography, give fidelities, $F$, in the range $0.91<F<0.96$ \cite{Rippe2008}. A brief description on the experimental setup, including the pulse-shaping system, will be given in Section~\ref{Experiment}, while Section~\ref{Initialization} describes how to efficiently initialize the system into well defined qubits, which are inhomogeneous subgroups out of these $10^{15}$ ions. In Section~\ref{Gates} we show experimental data on qubits prepared in arbitrary superposition states and characterized by quantum state tomography. The high fidelity results are all based on qubit operations that are robust against individual differences between the ions \cite{Roos2004,Rippe2005}. We will also present data from the development and use of optimal pulses sequences \cite{Timoney2008} for operation fidelities $>0.96$. Section~\ref{SingleIon} starts with a description of how to, in contrast to the experiments mentioned above, create qubits in these crystals that each consist of just a single ion, which offers a better scalability. The single ion qubit scheme \cite{Wesenberg2007} is based on co-doping the crystals with "readout ions". Such readout ions should have a short upper state lifetime in order to enable repeated cycling. One such candidate is the 4f to 5d transition in Ce$^{3+}$ and experimental data to characterize this transition is also presented in Section~\ref{SingleIon}. The last Section before the summary contains a discussion on entanglement of remote qubits and further scalability.

\section{Rare-earth-ion-doped crystals}
\label{RareEarth}
A major strength of using rare-earth ions as qubits are the long coherence times. In these ions, transitions within unfilled electron shells are partially shielded from environmental dephasing mechanisms by energetically lower, but outer-lying, filled shells. The present results were obtained using the $^3H_4 \rightarrow ^1D_2$ transition in \PrYSO at 606 nm \cite{Equall1995}. For temperatures below 4 K, the \Pr3 ions in this material has an electronic state lifetime of 164 \micro s, and a coherence time, \T2, of about 100 \micro s (somewhat depending on the density of excited states in the material). By applying magnetic fields however, this time can be increased, and there are also other rare-earth ions which have even longer coherence times, for example Er, where a \T2 of 6.4 ms has been measured \cite{Sun2002}.

\begin{floatingfigure}{45mm}
  \resizebox{45mm}{!}{\includegraphics{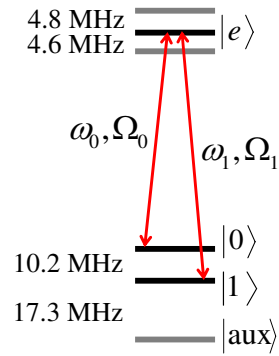}}
  \caption{Relevant energy level structure for the \Pr3 ion transition utilized in the present work.}
  \label{level_diag}
\end{floatingfigure}

The qubit states are not represented by electronically excited states though, but rather by ground state hyperfine levels, shown in the level diagram of \Pr3 in Figure~\ref{level_diag}. For \PrYSO, the hyperfine lifetime is about 90 s and the coherence time about 500 \micro s. The hyperfine \T2 can also be increased by magnetic fields, and up to 860 ms has been demonstrated \cite{Fraval2004}. In addition, dynamic decoupling sequences (bang-bang pulses) can also be applied to further increase the coherence time to more than 30~s~\cite{Fraval2005}.

The rare-earth ions are randomly doped into the host crystals, replacing one of the host ions in the lattice. This replacement causes changes to the crystal electric field which then causes each rare-earth ion to pick up a shift in its resonance frequency. This amounts to a large inhomogeneous broadening of the absorption line. For \Pr3, the inhomogeneous linewidth is about 5 GHz, which can be compared to the homogeneous linewidth of about 3 kHz, given by the optical \T2, mentioned above.

In the rare-earth quantum computing scheme, the inhomogeneous structure is of significant importance, as multiple qubits are addressed in the frequency domain, each qubit having a different resonance frequency. From above, the ratio of inhomogeneous over homogeneous linewidth is about $10^6$ for \PrYSO, but since ions do not have only a single transition frequency, the number of frequency channels are in practice much smaller. Fig.~\ref{level_diag} shows that the lowest and highest transition frequency for a given ion differ by $4.8+4.6+10.2+17.3=36.9$ MHz. Thus, about 100 individual frequency channels are a more realistic value. The \Pr3 ion sits in a non-centrosymmetric site, and thus has a permanent electric dipole moment. The value of this dipole moment depends on the electron configuration and the value of the dipole moment of an electronically excited state, is therefore different from that of the ground state. Qubit-qubit gates can then be executed through dipole-dipole interactions. Since the dipole moment changes when the ion becomes excited, exciting an ion will shift the transition frequency of all near-lying ions. An ion belonging to one qubit can then be controlled by the excitation of a another ion, belonging to another qubit, if they are spatially sufficiently close to each other. This effect is used as the control mechanism for conditional gates, and the basis for this was demonstrated in~\cite{Rippe2005}.

\section{Experimental setup}
\label{Experiment}
The experimental system is based around a Coherent 699-21 ring dye laser, stabilized to $~1$ kHz linewidth, by locking to a spectral hole in a \PrYSO crystal \cite{Julsgaard2007}. The rare-earth crystal used in the experiments is a 0.5 mm thick \PrYSO crystal with a \Pr3 doping concentration of 0.05\%, which gives a maximum optical density of about $2-3$. In order to get rid of phonon interactions, the crystal is submerged in liquid helium and cooled to 2 K. The available laser power is in the order of 100 mW before the cryostat, which translate into maximum Rabi frequencies of about 1-3 MHz, depending on which transition is targeted, given a laser focus of 100 \micro m.

High fidelity gates requires a high-performance pulse-shaping system, as will be further discussed in Section~\ref{Gates}. Two acousto-optic modulators (AOMs) are used to create the desired pulses. The incoming continuous wave (CW) photons are deflected by an RF acoustic wave in the AOMs. By varying the power of the acoustic wave, the amplitude of the deflected light is controlled. By virtue of energy and momentum conservation, the deflected photons must increase their frequency by an amount equal to that of the phonons. Thus, by varying the frequency of the acoustic wave we directly control the frequency of the light pulse, and furthermore, by changing the phase of the acoustic wave we also gain control over the optical phase of the light, which is necessary for controlling the coherent interaction with the ions. For these modulators, the power conversion from RF wave to light amplitude is quite non-linear. In order to achieve high accuracy of the pulse amplitude envelopes over as much as six orders of magnitude in dynamic range, an elaborate calibration system, that also includes the power dependence on RF frequency, was setup. The first AOM has a center frequency of 200 MHz and is aligned in a double pass configuration, in order to cancel out spatial movement of the deflected beam. After both passes the AOM has a total deflection range of 200 MHz. The second AOM has a 360 MHz center frequency, and is used only to split a pulse at a single frequency into two frequencies, such that an excitation pulse can be resonant with the $\mket{0} \rightarrow \mket{e}$ and $\mket{1} \rightarrow \mket{e}$ transitions at the same time, as further discussed in Section~\ref{Dark_gates}.

\section{Qubit initialization}
\label{Initialization}
\begin{figure}
    \begin{center}
        \includegraphics[width=440pt]{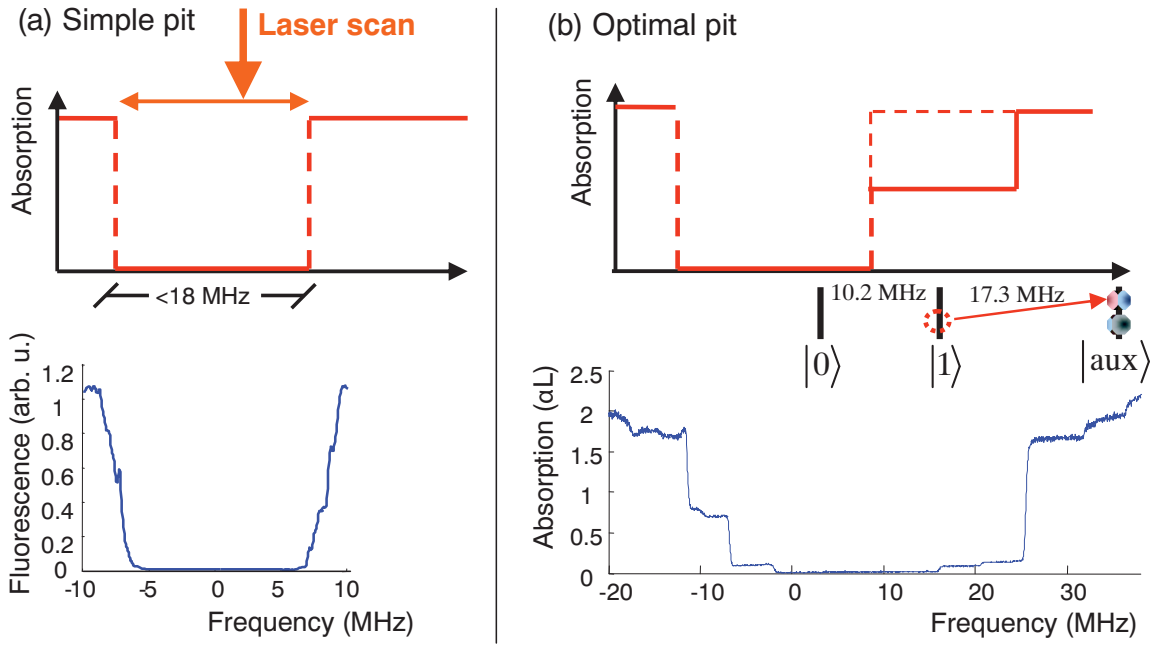}
        \caption{(color online) a) shows a simple pit created by repeated scanning across an interval less than 18 MHz. Panel b) displays an improved pit, where scans on multiple intervals have been iteratively repeated in order to shuffle all ions outside the pit as far out as possible, thereby creating an optimal pit. Also note that the scales in the two experimental figures are different.}
    \label{Pit_creation}
    \end{center}
\end{figure}
Starting from the inhomogeneous distribution, we employ advanced optical hole-burning sequences in order to extract a useful qubit system. Initially, the ions occupy all three ground states, and the absorption at any given frequency within the absorption profile, contains contributions from all nine possible transitions (see Figure~\ref{level_diag}). The goal of the hole-burning is to initialize a narrow controlled ensemble of ions into the \ket{0}-state, well separated in frequency from any other ion.

At first, a wide spectral hole, henceforth called a \emph{pit}, is created. The hyperfine level splittings set the limit for the maximum frequency range of such a pit. Since the ions must be in one of the ground states, it can not be larger than 27.5 MHz, as can be seen in Figure~\ref{level_diag}. This maximum width is then further reduced by the excited state splitting down to 18.1 MHz. Scanned laser pulses target specific transitions inside this $~$18 MHz region, and are repeated in order to create a spectral pit where all ions have been removed within an interval of maximum 18 MHz. Such a simple pit is shown in Figure~\ref{Pit_creation}a.

Eventually, qubit operation pulses will be used inside the pit structure, and it then becomes important to minimize the interactions between the tails of such pulses and the edges of the pit. The pit shown in Figure~\ref{Pit_creation}a displays non-optimal edges, that are a bit rounded and also go up to the maximum \al immediately. In order to improve on this, a more elaborate pit-burning scheme was introduced. In this scheme, pulses that target specific levels of \Pr3 outside the pit are used in order to burn them even further out, as illustrated by the top part of Figure~\ref{Pit_creation}b. For an ion having a particular frequency relative to the pit, as the ion illustrated in this figure, level \ket{0} and \ket{1} are targeted in an optical pumping process that eventually transfers them to their \ket{aux} state. After this, ions of other frequency ranges might fall back down inside the pit, and the initial pit-burning pulses will have to be repeated iteratively together with the secondary pulses to keep the center of the pit empty. A similar procedure is then also repeated on the left side of the pit in order to achieve an optimal result, shown in the lower part of panel b. A more extensive description of the burning procedure can be found in \cite{Guillot-Noel2009} and a detailed list of the exact pulses used to burn such an optimal pit can be found in \cite{Walther2009}.

\begin{figure}
    \begin{center}
        \includegraphics[width=400pt]{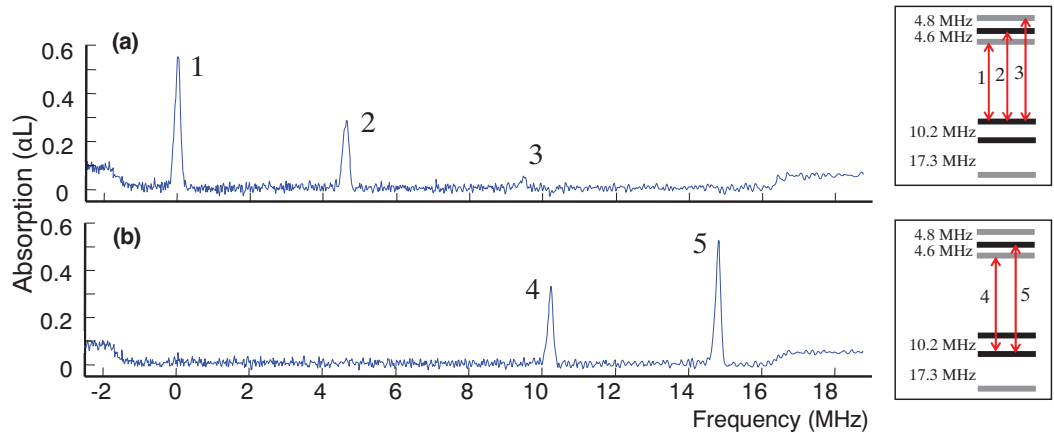}
        \caption{(color online) a) shows the rare-earth qubit system initialized to the \ket{0} state, visible through the three absorption peaks from this state. Panel b) then shows the system after a coherent transfer over to the \ket{1} state.}
    \label{peaks}
    \end{center}
\end{figure}

After an empty spectral pit has been created, a qubit ensemble can be initialized inside the pit by a \emph{"burnback"} pulse sequence. These pulses can with benefits be coherent pulses that transfer ions from outside the pit, say at higher frequencies, first to the excited state, and from there down to one of the ground states in a controlled manner. Figure~\ref{peaks}a shows the system initialized to one of the ground states, in this case the \ket{0} state.

\section{Qubit gate operations}
\label{Gates}
\begin{figure}
    \begin{center}
        \includegraphics[width=440pt]{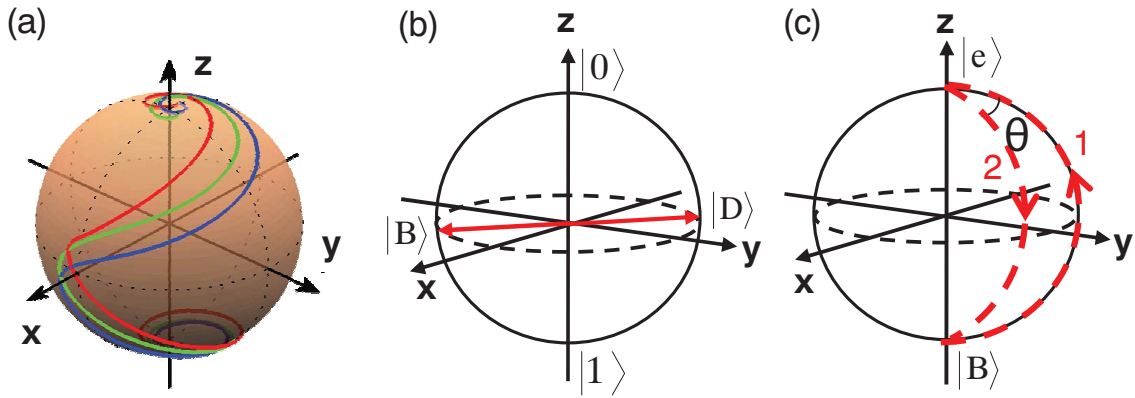}
        \caption{(color online) a) shows the paths taken by three ions with different resonance frequencies, while being transferred to the excited state by a coherent sechyp pulse. Panel b) illustrates the dark and the bright states on the Bloch sphere, relevant for the gate operations, and panel c) shows a full qubit gate operation by means of two consecutive dark state pulses (two-color pulses). Note that the transfer in c) is going from and to the \emph{bright} state.}
    \label{dark_state_path}
    \end{center}
\end{figure}

The ensemble qubit is now initialized to the \ket{0} state. In order to be able to perform high fidelity gate operations, we must now find pulses that compensate for the inhomogeneous broadening of the ensemble, which would otherwise cause dephasing during the excitation. One type of pulses that are robust against such dephasing are the complex hyperbolic secant pulses, or \emph{sechyp} for short. The amplitude envelope of these follow a hyperbolic secant shape, while at the same time the pulse is chirped according to a hyperbolic tangent function, centered around the mean ensemble transition frequency. Such pulses can efficiently transfer an inhomogeneous ensemble between the two poles on a Bloch sphere, as illustrated in Figure~\ref{dark_state_path}a. The different lines correspond to ions with different atomic resonance frequencies, which can all be seen to converge at the top. The robustness of these pulses has been verified experimentally by transfer efficiencies of $~97\%$ for ensembles inhomogeneously broadened more than 100 linewidths. Figure~\ref{peaks}b shows the system after such a sechyp transfer from $\mket{0} \rightarrow \mket{e}$ immediately followed by a transfer from $\mket{e} \rightarrow \mket{1}$. In addition to this, the sechyp pulses also have the good property that they are robust against fluctuations in laser power. While normal square- or gaussian-shaped pulses need to have very well tuned values of laser intensity and duration in order to get the correct pulse area, the sechyp pulses simply need to be above a certain threshold Rabi frequency, above which they always work with high efficiency. This independence of the light intensity can be intuitively perceived from the Bloch sphere in Figure~\ref{dark_state_path}a, where we see that as the different components approach their final state, they circle around the top pole, and a higher Rabi frequency would just make them circle around the z-axis faster, not take them away from the desired state. These pulses were investigated more carefully in~\cite{Roos2004}.

\subsection{Dark state gates}
\label{Dark_gates}
The sechyp pulse-shapes solve the problem of inhomogeneous dephasing as well as offer a robustness with respect to fluctuating experimental parameters, but can only perform pole-to-pole transfers, not create arbitrary superposition states. The qubit is the two hyperfine levels denoted \ket{0} and \ket{1} in Figure~\ref{level_diag}, and to perform operations between these states, two-color fields, which are resonant with $\mket{0} \rightarrow \mket{e}$ and $\mket{1} \rightarrow \mket{e}$ at the same time, are employed \cite{Roos2004}. Determined by the phase relation, $\phi$, between the two fields, two superposition states, shown in Figure~\ref{dark_state_path}b, the bright and dark state:
\eq{eq:dark_basis}{
\left\{ \begin{array}{ll}
  \mket{B} = \frac{1}{\sqrt{2}}\left(\mket{0}-e^{-i\phi}\mket{1}\right)\\
  \mket{D} = \frac{1}{\sqrt{2}}\left(\mket{0}+e^{-i\phi}\mket{1}\right)
\end{array} \right.
}
can be defined. Similarly to electromagnetically induced transparency (EIT), the dark state does not interact with the light. Thus, the two-color pulse will only target the bright state, and two consecutive transfers, first from $\mket{B} \rightarrow \mket{e}$ and then from $\mket{e} \rightarrow \mket{B}$, but going down along a different path as illustrated in Figure~\ref{dark_state_path}c, will add a phase angle enclosed by the rotation, $\theta$, to the bright state: $\mket{B} \rightarrow e^{i\theta}\mket{B}$. This operation can be rewritten in the computational basis (\ket{0},\ket{1}), and then becomes
\eq{eq:qubit_rot}{
  U_{dark} = e^{i\theta/2}\left[\begin{array}{cc}
  \cos(\theta/2) & ie^{i\phi}\sin(\theta/2)\\
  ie^{-i\phi}\sin(\theta/2) & \cos(\theta/2)
  \end{array} \right].
}
which is a matrix describing an arbitrary rotation around any vector on the equator of the Bloch sphere. By picking suitable phase angles, one can thus perform any qubit gate operation \cite{Roos2004}.

\begin{figure}
    \begin{center}
        \includegraphics[width=400pt]{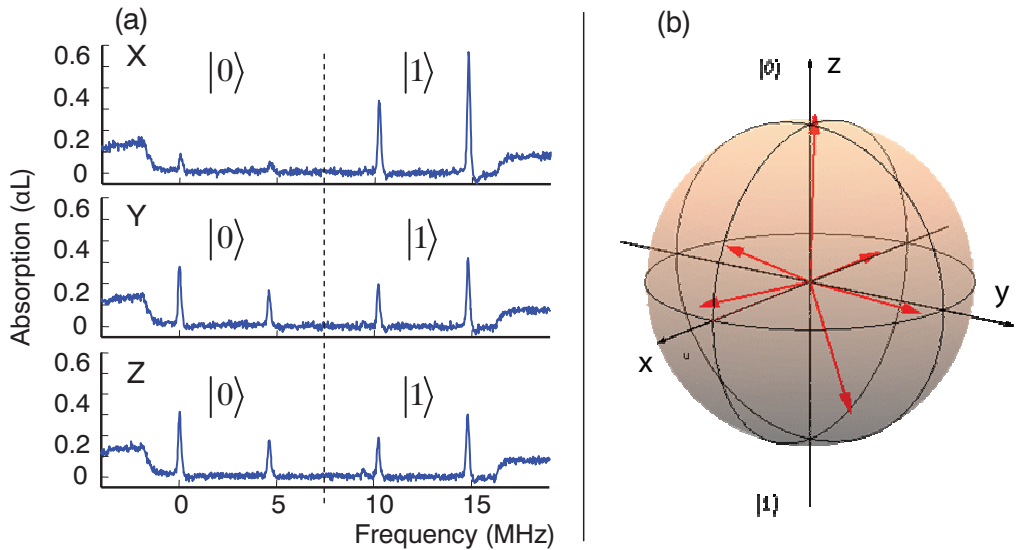}
        \caption{(color online) a) shows the projective measurements on the $x$-, $y$- and $z$-axis respectively. The two left peaks are both transitions from the \ket{0} state and correspond to a $+1$ measurement value, while the two right peaks similarly are transitions from the \ket{1} state and correspond to a $-1$ value with the respect to the current axis. Panel b) shows the result of a quantum state tomography of the six states along the positive and negative axes.}
    \label{QST}
    \end{center}
\end{figure}

These dark state gates were implemented and the qubit superposition states that were created as a result of the rotations, were characterized using quantum state tomography. In Figure~\ref{QST}a the readout of a particular superposition state, $\frac{1}{\sqrt{2}}\left(\mket{0}-\mket{1}\right)$ is shown as an example. The three curves correspond to projection measurements on the $x$-, $y$- and $z$-axis respectively. As expected, we see that both the $y$ and the $z$ measurement contain equal contributions from either state, but the $x$ measurement gives a projection onto one of the states. States corresponding to the six different (positive and negative) axes on the Bloch sphere, were prepared, and characterized, and the resulting state vectors inscribed in a Bloch sphere are displayed in Figure~\ref{QST}b.

The $z$-projection was done through an absorption measurement, where the two leftmost peaks in Figure~\ref{QST} correspond to transitions from the \ket{0} state and the two rightmost peaks correspond to transitions from the \ket{1} state. In order to do the $x$- and $y$-measurements, suitable rotations were first performed to rotate those axes onto the $z$-axis, such that an absorption measurement now gives the same result as a direct $x$- or $y$-measurement would have, in the original basis. From those measurements the density matrix can then be calculated through (see e.g. \cite{Nielsen2000})
\eq{eq:qubit_density}{
  \rho = \frac{\mathrm{tr}(\rho)I+\mathrm{tr}(X\rho)X + \mathrm{tr}(Y\rho)Y + \mathrm{tr}(Z\rho)Z}{2},
}
where $X$, $Y$ and $Z$ correspond to the Pauli matrices, which together with the identity, $I$, span the $2\times2$ qubit space in the density matrix representation. The above equation is useful because the traces correspond to actual physical measurements. For example, $\mathrm{tr}(Z\rho)$ means measuring the projection of the unknown state onto the $Z$ axis, which will yield a value between -1 and +1, and the same for the other axes. The fidelity of the entire procedure can then be calculated from
\eq{eq:fidelity}{
  F_{tot} = \left<\psi_{theor}\left|\rho_{exp}\right|\psi_{theor}\right>,
}
i.e. the overlap between the theoretical and the experimental representations. For the six states displayed in Figure~\ref{QST}b, the total fidelities were all between 0.84 and 0.92. As mentioned above the total procedure consists of two rotations, first one to create the state, then another one to perform the measurement in the right basis. Thus, the fidelity for a single qubit gate can be expressed as $F_{gate} = \sqrt{F_{tot}}$, and we see that the qubit gate fidelities are in the range of $0.91<F_{gate}<0.96$. More information and discussion regarding these results can be found in \cite{Rippe2008,Walther2009}.

\subsection{Optimal control pulses}
Even though the dark state qubit gates were realized with good fidelity there are still reasons to search for further improvements. The total duration of the dark state pulses that performs one arbitrary qubit rotation is close to 18 \micro s. The employed laser power on the other hand, typically provides Rabi frequencies in the order of 1 MHz, which indicates that the qubit rotations could probably be performed an order of magnitude faster, with an optimal scheme. To reduce the operation time, pulses obtained through optimal control theory calculations are currently investigated.

\begin{figure}
    \begin{center}
        \includegraphics[width=440pt]{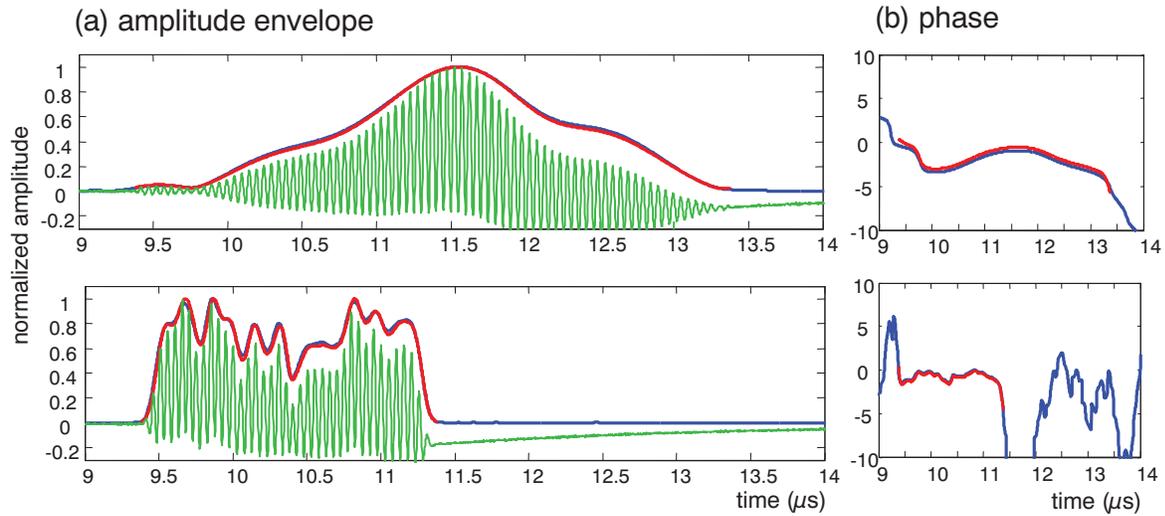}
        \caption{(color online) Two pulse shapes with different bandwidths, both designed by optimal control algorithms to perform efficient state-to-state transfers. The left side shows the detected beating and the amplitude envelope, while the right side shows the phase evolution. In both cases the red line is the theoretically desired shape, while the blue line is the experimentally detected shape, and in both cases they are matching very well. Note that in the phase plot, the phase does not have any meaning outside the pulse envelope (since the amplitude there is zero).}
    \label{opt_pulses}
    \end{center}
\end{figure}

Optimal control theory provides a framework for finding the best set of controls to steer a system so that a desired target state or unitary gate is implemented. In some special cases these controls can be
determined analytically, see for example \cite{Khaneja2001,Khaneja2003,Fisher2009}. In addition, powerful numerical methods \cite{Khaneja2005} are available that make it possible to explore the physical limits of time-optimal control experiments in cases where no analytical solutions are known. These methods have so far been successfully applied to spin systems \cite{Skinner2003,Kobzar2004,Kobzar2005} and superconducting qubits \cite{Sporl2007a}, and trapped ions \cite{Nebendahl2009}. Optimized pulses can be designed to account for experimental errors, and, if a realistic model is available, can include multiple system levels and transitions.

The basic idea of optimal control algorithms is to minimize a cost function, which is determined by the physical system at hand. In the present case, this is given by the light-field interaction with the \Pr3 ion, and in addition to this, a number of experimental limitations, such as AOM maximum bandwidth or laser power fluctuations, are enforced as well. The algorithm then finds a pulse shape that performs a particular operation to a specified minimum fidelity, while staying within the restrictions. It can be noted that the solution is not necessarily unique, or even at the global maximum, and the shapes that comes out of the calculations are then often strange-looking, without obvious intuitive explanations. In order to characterize how well the AOM system could reproduce these strange shapes, an interference experiment was set up, where the AOM pulse-shaping system is put in one arm and the light passing through this part then interferes with the light from an undisturbed part of the laser. The result is a beating, from which both the amplitude envelope and the phase evolution of the pulses can be obtained via a Fourier transform. Figure~\ref{opt_pulses} shows two examples of such pulses. In panel a) of this figure, the green, oscillating curve is the detected beating, the blue line is the amplitude envelope of the beating, which is almost perfectly overlapping with the red line, which is the theoretically desired shape (may be hard to see if not in color).

\begin{figure}
    \begin{center}
        \includegraphics[width=300pt]{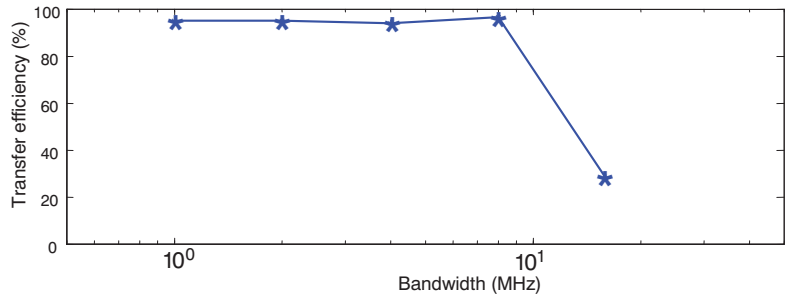}
        \caption{Transfer efficiency of $\mket{0} \rightarrow \mket{e}$ by means of optimal control pulses, as a function of bandwidth.}
    \label{opt_eff_vs_bw}
    \end{center}
\end{figure}

The upper and lower pulse differs in allowed bandwidth. The lower pulse has a bandwidth of about 16 MHz, while the upper pulse has been restricted to work with a bandwidth of only 2 MHz. Ideally, one would want as high bandwidth as possible, since this enables pulses with shorter duration. The right side of Figure~\ref{opt_pulses} displays the phase evolution of the detected pulses as well as the desired phase chirp, and again, there is almost perfect overlap. In order to test the optimal control results, a series of five pulses were designed, each one with the same objective, to do a state-to-state transfer from \ket{0} to \ket{e}, but with different bandwidths. The resulting efficiency as a function of bandwidth is shown in Figure~\ref{opt_eff_vs_bw}. As can be seen, the efficiency is very good for all bandwidths except the highest (16 MHz). The reason for this is not yet fully known, but is believed to be related to complications that arise when multiple levels are involved. From the level diagram (Figure~\ref{level_diag}) we see that any pulse that centers around the $\mket{0} \rightarrow \mket{e}$ transition and has a bandwidth less than $~2 \cdot 4.6 = 9.2$ MHz will only involve those two levels and nothing else, but any pulse with a higher bandwidth will automatically involve at least the two other excited states as well, which creates a much more complicated situation. For two-level schemes, other groups have also obtained very good results, for example in ion traps \cite{Timoney2008}. But clearly, there should also be suitable optimal pulses for three-level configurations. Thus, the good results for the simpler pulses together with the potential of an order of magnitude faster pulses make the optimal control technique seem promising for the future.

\section{Single-ion qubits}
\label{SingleIon}
\subsection{Scalability of the ensemble qubit approach}
Scaling to large number of qubits is a key issue for a quantum computer. In the present system, the qubit-qubit operation is achieved by the dipole-dipole interaction between two neighboring ions. As stated in Section~\ref{RareEarth} they must physically sit close enough to each other so that when the state of one ion is changed, e.g. from the ground state to the excited state and the electric field from the static dipole moment changes, this will shift the neighboring ion out of resonance with its original transition frequency. In this way, closely lying ions can control each other. For an ion absorbing at a selected frequency, $\nu_i$, there will be several ions in the vicinity that can control this ion. In the ensemble approach, for two qubits, $i$ and $j$, at transition frequencies $\nu_i$ and $\nu_j$, there is a probability, $p$, that for an arbitrary ion, $a_i$, in qubit $i$, there will be an ion in qubit $j$ that can control $a_i$. For $n$ qubits, if $N$ denotes the number of ions in a qubit that can be controlled by ions in the other $n-1$ qubits, we get that $N\propto p^{n-1}$. For the materials studied so far $p$ is in the order of 1\%, which means that for a 5-qubit quantum register there will only be 1 out of $10^8$ ions that are useful in each qubit.

Several remedies were pointed out to improve the scaling property by pushing $p>1$, e.g. choosing ions with large static dipole moment, increasing the dopant concentration \cite{Nilsson2004b}, constructing the qubit-qubit interaction by using a mediator bus ion \cite{Wesenberg2007} or adding a specially selected readout ion to detect the state of single ions, thereby enabling the use of single ions instead of an ensemble of ions for each qubit. Considering both advantages and disadvantages in each scheme the single ion readout approach is believed to be the easiest way to achieve the scalability.

\subsection{The readout-ion approach}
Inside any small volume within a rare-earth-ion doped crystal there is a high probability to find several strongly interacting ions which can control each other and each of them can represent one qubit. A readout scheme has been suggested for detecting the qubit state in such a single ion qubit system through the dipole-dipole interaction \cite{Wesenberg2007}. The dopant concentration of the readout ion should be very low to make sure there is on average only one readout ion interacting with the laser field. To fulfill its role, the readout ion is supposed to meet several requirements: (i) The absorption spectrum should be well separated from that of the qubit ion so that the readout procedure does not affect the qubit state. (ii) The excited state lifetime should be short and there should not be any trapping state. Thus, the excitation may be cycled many times producing many photons and giving a strong detection signal. (iii) The homogeneous linewidth should be narrow so that when a nearby qubit ion is transferred between the ground and excited states, the dipole-dipole interaction is able to shift the readout ion in or out of resonance with the readout beam frequency, which consequently turns on or off the fluorescence. (iv) The inhomogeneous linewidth should be large which provides a larger number of frequency channels.

\subsection{Find and search schemes}
\begin{figure}
    \begin{center}
        \includegraphics[width=300pt]{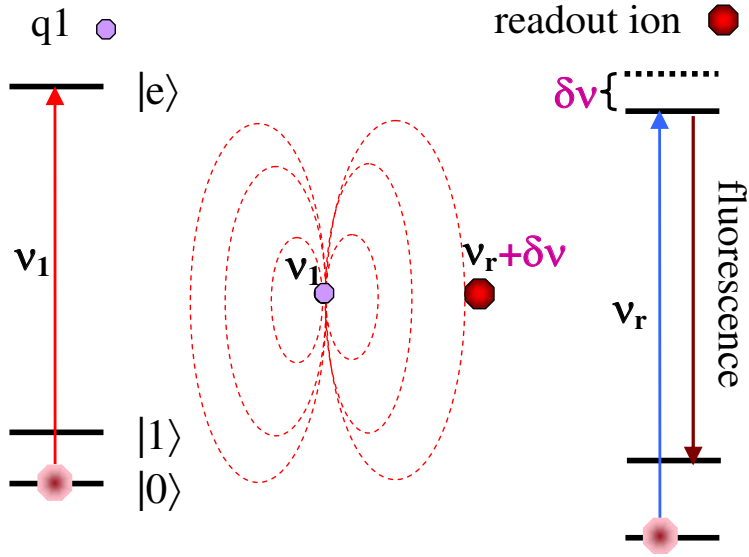}
        \caption{(color online) The dipole-dipole interaction between a qubit ion and a near-lying readout ion. If the qubit ion $q1$ is excited from the \ket{0} state to the \ket{e} state at frequency $\nu_1$, the electric field generated by the dipole will shift the transition frequency of the readout ion by an amount $\delta\nu$, after which excitation at frequency $\nu_r$ will have no effect and the readout ion fluorescence will be turned off.}
    \label{dipole_dipole_interaction}
    \end{center}
\end{figure}

Provided the spectral information of the readout ion is characterized, the nearest qubit, $q1$, and further qubit chain structures in the vicinity of the readout ion, can be mapped out using the fluorescence signal from the readout ion. After detecting the readout ion fluorescence at a certain excitation frequency, $\nu_r$, one would keep the readout laser working at $\nu_r$, while scanning the qubit laser across the inhomogeneous qubit transition frequency to find a frequency, $\nu_1$, where the fluorescence signal stops. This would mean that a qubit ion with the addressing frequency $\nu_1$ can control the readout ion, as illustrated in Figure~\ref{dipole_dipole_interaction}. With the qubit and readout laser working at $\nu_1$ and $\nu_r$ respectively, the presence or absence of fluorescence tells whether $q1$ is in the \ket{1} or the \ket{0} state. The general case is that the qubit is in a superposition state, then one has to do the quantum state tomography measurements as described in Section~\ref{Dark_gates}, then execute the above readout procedure to characterize the probability distribution between the \ket{0} and \ket{1} state for each projection. After the excitation frequency of $q1$ is known one can then extend the readout scheme to find out which qubit, $q2$, can control $q1$ through the qubit-qubit interaction and further, consecutively map out the whole interacting qubit chain and read out all qubits.

\subsection{Investigation for a readout ion}
So far $\mathrm{Ce^{3+}}$ is believed to be the most promising candidate for the readout ion. The excitation wavelength for the $4\mathrm{f}-5\mathrm{d}$ zero-phonon-line (ZPL) of $\mathrm{Ce^{3+}}$ ion randomly doped in the YSO-crystal is around 371 nm, which is well separated from the qubit ion transition frequency 606 nm (Pr) or 580 nm (Eu). The $\mathrm{Ce^{3+}}$ homogeneous linewidth is determined by detecting the sum frequency signal from a saturation spectroscopy experiment with different amplitude modulation frequencies on the pump and probe beams. The measured homogeneous linewidth was as narrow as 3 MHz. This is a very encouraging result, because based on a reasonable assumption about the Ce dipole moment, the Pr ion should be able to shift the Ce transition line by 30 MHz when they are separated by 7 nm \cite{Ahlholm2008}. With a 50 ns lifetime \cite{Hernandez2006,Aitasalo2004} this also means that the homogeneous linewidth is lifetime-limited which is the ideal case for this readout ion concept.

The inhomogeneous linewidth of the ZPL in a crystal with 0.08\% cerium concentration is measured to be 80GHz at 4K. However, the fluorescence at the center of the ZPL is only twice the background signal outside the ZPL. The exact origin of this background is unknown, but it could for example be phonon sideband transitions, which are stronger for $4\mathrm{f}-5\mathrm{d}$ transitions than for $4\mathrm{f}-4\mathrm{f}$ transitions. Clearly this will affect the readout fidelity. Assume the detector quantum efficiency is 10\% and the fluorescence collection efficiency is 30\% \cite{Palm1999}, and a $\pi$-pulse is applied on the $\mket{0} \rightarrow \mket{e}$ qubit transition. As the readout laser is turned on, there will then be about 100 fluorescence photons detected during the optical lifetime (150 \micro s) of the \Pr3 ion, if the qubit was in its \ket{1} state. On the other hand, if the \Pr3 ion was in its \ket{0} state, the ZPL background fluorescence might still give around 50 photons due to off-resonant excitations (the background signal). However, the difference in the number of photons is certainly still sufficient for separating the two cases. One way to reduce the background signal, would be to decrease the number of Ce ions within the laser focus.

%%% old section
%On the other hand, we have also detected a background signal far outside the homogeneous linewidth which appears to come from off-resonant excitation of Ce ions. This could for example be phonon sideband transitions, which are stronger in $4\mathrm{f}-5\mathrm{d}$ transition than in $4\mathrm{f}-f\mathrm{f}$ transitions. To eliminate this background signal, which would improve the signal to noise ratio, one can reduce the excitation volume either by using a single layer doped Ce:YSO crystal or by using an evanescent field for the readout measurement.
%To illustrate the effects of the observed background, we may note that the inhomogeneous linewidth of the ZPL in a crystal with 0.008\% cerium concentration is measured to be 80{GHz} at 4K. The fluorescence at the center of the ZPL is 3 times higher than the background signal outside the ZPL. Assume the quantum efficiency is 10\% and the fluorescence collection efficiency is 30\% \cite{Palm1999}, and a $\pi$-pulse is applied on the $\mket{0} \rightarrow \mket{e}$ qubit transition. Turning on the readout laser, there will then be about 100 fluorescence photons detected during the optical lifetime (150 \micro s) of the Pr ion if the qubit was in its \ket{1} state. On the other hand if the Pr ion was in its \ket{0} state, the ZPL background fluorescence would still give around 30 photons due to off-resonant excitations. However, this difference in the number of photons is certainly sufficient for separating the two cases.

\section{Scalability}
\label{Scalability}
When the single-ion readout system is scaled up by adding new qubits to the (possibly branched) qubit ion chain there will, although there are many frequency channels available within the inhomogeneously broadened qubit transition line, eventually be a new qubit ion (\mrm{Q_{new}}) that have the same frequency as one of the qubit ions (\mrm{Q_{old}}) already in the chain. In this situation, the two ions (\mrm{Q_{old}} and \mrm{Q_{new}}) cannot any longer be individually addressed in frequency space. It is conceivable that there are algorithmic solutions to this problem, where single qubit operations could be replaced by two-qubit operations, so that operations are applied to "the qubit with frequency $\nu_i$, sitting next to a qubit with frequency $\nu_j$". As an alternative, the problem of coinciding transition frequencies can be removed by mounting (closely spaced) electrodes onto the crystal. The part of the qubit chain on which the operations are carried out, is selected by applying a voltage on the appropriate electrode. Eu in Y$_2$SiO$_5$ has a Stark coefficient of 35 kHz/(V/cm) \cite{Graf1998}, and ions close to the electrode can be shifted into resonance with the excitation pulse. A 1 MV/cm field would shift the transition frequency 35 GHz, a detuning much larger than the inhomogeneous transition line width. Simulations show that 20 nm long electrodes separated by 40 nm on a surface may give shifts of the order of 30 GHz within a 60 nm region while ions 20 nm further away would shift less than 10 GHz \cite{Wesenberg2007}. However, still smaller electrodes and electrode separations would be favorable and while this is in line with the current technological development we will also now analyze the possibility of entangling spatially remote few-qubit systems.

Thus, we consider spatially separated qubit chains, each containing a not too large number of qubits. There are several schemes for entangling spatially remote few-qubit systems. To optically entangle a qubit in one system with a qubit in another system the key point is to configure the photon detection (or absorption) process such that when a photon is detected (absorbed) it cannot in principle be determined which of the two systems that emitted (absorbed) the photon \cite{Duan2003,Barett2005}. If the set-up is appropriately configured, the photon detection/absorption entangles the two qubits.

Scalable quantum computing is then achievable also with errors in the remote entanglement in the order of several percent, provided that the error probability for operation within each of the qubit chain is below $10^{-4}$ \cite{Jiang2007}. Remote entanglement processes often correspond to a probabilistic event, which is repeated until it succeeds. Clearly, schemes with very high probability to succeed in a single try are preferable. Such schemes can for example be devised using optical micro-cavities \cite{Cirac1997}, e.g., toroidal micro-resonators \cite{Vahala2003,Spillane2005} and can also be improved by the use of cluster states \cite{Raussendorf2001}.

We consider a rare earth crystal with a Ce readout ion, close to the surface facing a toroidal microcavity. Around this readout ion there will be a number of qubit ions that interact and can be read out, as described in Section~\ref{SingleIon}. One of these qubit ions is selected as the \emph{entangler ion}. The purpose is now to entangle \emph{entangler ions} in different rare-earth crystals. Such a scheme requires that also rare-earth ions close to the crystal surface retain their favorable coherence properties, and preliminary data indicate that this indeed is the case \cite{Longdell2008}.

\begin{figure}
    \begin{center}
        \includegraphics[width=400pt]{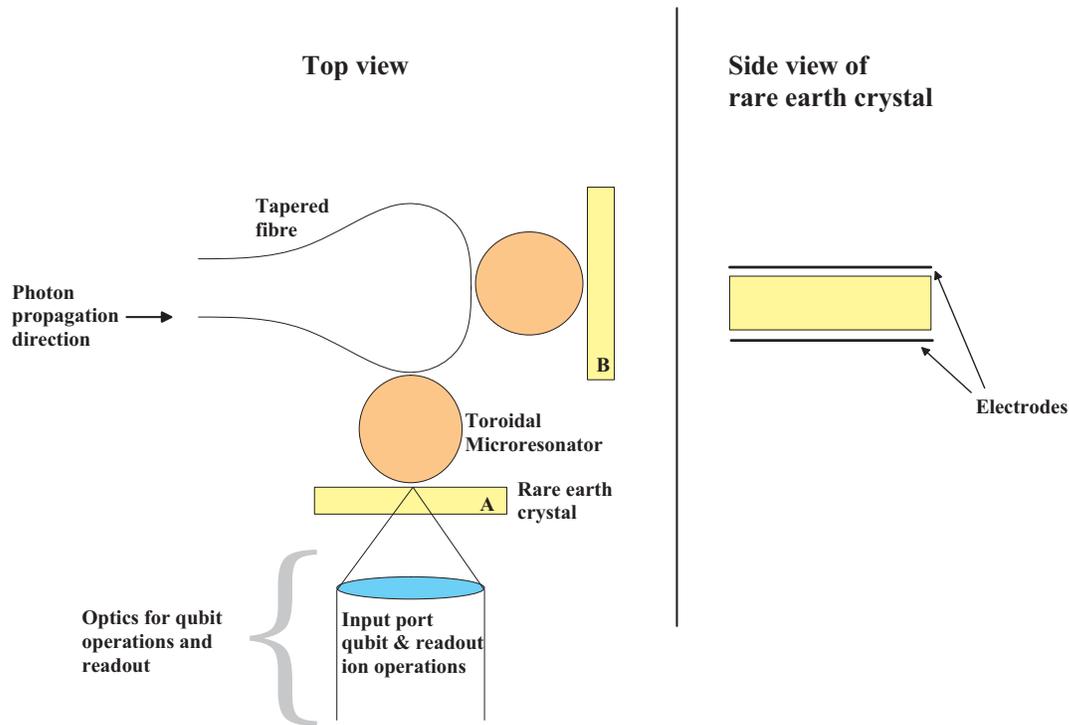}
        \caption{(color online) Schematic view of an approach to entangle spatially separate few qubit processors.}
    \label{scalableqc}
    \end{center}
\end{figure}

Consider Figure~\ref{scalableqc}, the interaction between the cavity and the ion is adjusted such that a single photon corresponds to a $\pi$-pulse \cite{Xiao2004}. Using tapered fibres, a single photon can be coupled into the toroidal microcavity with an efficiency (ideality) $>99.98\%$ \cite{Spillane2003}. The entangler ion in the first crystal, A, is prepared in the superposition state $(\mket{0}+\mket{1})_A$ and the entangler ion in the second crystal, B, is prepared in state \ket{1}$_B$. As experimentally demonstrated \cite{Macfarlane1994}, the ion transition frequencies to an excited state, \ket{e}, can readily be tuned by an electric field. Thus, by tuning the $\mket{0} \rightarrow \mket{e}$ transition for the ion in A to the same frequency as the $\mket{1} \rightarrow \mket{e}$ transition for the ion in B using electric fields (see right hand side of Figure~\ref{scalableqc}) and sending a photon in the tapered fibre at this frequency in the direction indicated in the figure, the two ions will be prepared in the state ($\mket{e}_A\mket{1}_B + \mket{1}_A\mket{e}_B$). Applying local $\pi$-pulses on each of $\mket{e}_A \rightarrow \mket{0}_A$ and $\mket{e}_B \rightarrow \mket{0}$ at the respective input ports for the qubit and the readout ion, then creates the Bell state $\mket{01} + \mket{10}$.

In a more complete version of the approach in Figure~\ref{scalableqc}, there would be a large number of microresonators (with their corresponding crystals) along the tapered fibre. Which two crystals that are selected for any specific entanglement process can be controlled by e.g. adjusting the distance between the tapered fibre and the resonator (cavity) using e.g. micro-actuators. Alternatively, the cavity can be tuned off resonance with the transition and photon frequency. Tuning the cavity can be carried out using expansion due to temperature. For example, electrodes can be connected to the toroid and the substrate. The temperature and thereby also the frequency change will then be proportional to the dissipated power \cite{Armani2004}. In general the system could be quite flexible and versatile, as an example, ions can be tuned in and out of resonance with the cavity using an electric field across the crystal.

As earlier stated, the qubit coherence times can be 30 seconds for Pr ions \cite{Fraval2005} and for Eu ions coherence times of the order of an hour have been predicted \cite{Longdell2006}. A strength with the scheme outlined above is therefore that there is ample time both to create entanglement and to carry out operations once the systems have been entangled.

To compare with some existing data, free Cs atoms can have a single photon Rabi frequency of 100 MHz for Cs atoms 45 nm from the toroidal surface \cite{Aoki2006}. For \PrYSO which has an oscillator strength almost six orders of magnitude lower than free Cs atoms \cite{Graf1998}, would then yield a Rabi frequency of about 150 kHz. This value is too low for a single photon to act as a $\pi$-pulse and we are therefore looking at techniques for enhancing the interaction between an ion just below the surface and an evanescent field.

\section{Concluding remarks}
\label{Summary}
To summarize, single-qubit operation fidelities above 90\% are obtained for ensemble qubits and $>99\%$ single qubit operation fidelity should be readily obtained with single-ion qubits. However, the single ion qubit scheme is contingent upon the ability to read out the hyperfine state of single rare-earth ions and work to develop this capacity is still ongoing. Although, based on current data in the investigation of using Ce$^{3+}$ as a readout ion for such a scheme, there is reason to be quite optimistic about the possibilities to succeed. Finally, it may be pointed out that while quantum state storage in rare-earth-ion-doped materials is pursued vigorously by several groups and now demonstrates excellent progress \cite{Riedmatten2008} including storage efficiencies above 30\% \cite{Amari2009}, only two groups \cite{Longdell2004, Rippe2008} have published experimental efforts on quantum computing in rare earth crystals and considering the limited efforts, substantial progress has been made.

\section*{Acknowledgements}
This work was supported by the Swedish Research Council, the Knut and Alice Wallenberg Foundation, the European Commission through the integrated project QAP and the Crafoord Foundation.\\

% Bibliography
\bibliographystyle{unsrt}
\bibliography{../bibtex/Ref_lib}

\end{document}